\documentclass[10pt]{article}
\usepackage{amssymb}
\input epsf
\begin{document}
\title{
\hfill{}\\
\hfill{}\vspace*{0.5cm}\\
\sc
Comments on ``Differential cross section for Aharonov-Bohm effect
with nonstandard boundary conditions''
\vspace*{0.3cm}}
\author{ {\sc Alexander Moroz}
\vspace*{0.3cm}}
\date{
\protect\normalsize
\it FOM-Instituut voor  Atoom- en Molecuulfysica, Kruislaan 407, 
1098 SJ Amsterdam, 
The Netherlands}
\maketitle
\begin{center}
{\large\sc abstract}
\end{center}
We show that the violation  of rotational symmetry
for differential cross section for Aharonov-Bohm effect
with nonstandard boundary conditions has been known 
for some time. Moreover, the results were applied to discuss
the Hall effect and persistent currents of fermions 
in a plane pierced by a flux tube.

\vspace*{0.6cm}


\vspace*{1.9cm}

\begin{center}
({\bf Europhys. Lett. 47, 273-274 (1999)})
\end{center}

\thispagestyle{empty}
\baselineskip 20pt
\newpage
\setcounter{page}{1}
\noindent In their recent paper \cite{SV}, \v{S}\v{t}ov\'{\i}\v{c}ek and 
V\'{a}\v{n}a discuss the differential cross section $d\sigma/d\theta$
for the Aharonov-Bohm effect with nonstandard boundary conditions.
The latter mean in general that a wave function does not vanish
at the position of a singular flux tube.
Their main result is that, compared with the standard case, two new 
features occur: a violation  of rotational symmetry and a more 
significant backward scattering. They discuss the violation  of 
rotational symmetry in the rotationally invariant case 
(the case $w'=0$ in \cite{SV}) and for a new class 
of self-adjoint extensions which do not commute with the 
angular momentum (see also \cite{KS}). In the latter case, the scattering
cross section depends on the incident direction $\theta_0$ and the 
outgoing direction $\theta$ separately and is not invariant
under $\theta_0,\, \theta \rightarrow \theta_0+\gamma,\, \theta+\gamma$.
In the rotationally invariant case, the scattering
cross section depends on the scattering angle
$\varphi=\theta -\theta_0$, nevertheless, 
$d\sigma(\varphi)/d\theta\neq d\sigma(-\varphi)/d\theta$ in general.

We should like to point out that 
the discussion  of the scattering in the presence of an Aharonov-Bohm 
potential  with the nonstandard boundary conditions
has a much longer history (see, for example, \cite{PG})
than references in \cite{SV} tend to indicate.
Also the violation  of rotational symmetry 
for the differential cross section in the rotationally invariant case
has been pointed out some time ago \cite{AM}. 
In Ref. \cite{AM}, various physical quantities were calculated
for a two-parameter family of self-adjoint extensions which respect
the rotational invariance (i.e., no coupling between different 
angular momentum channels), including the local density of states,
persistent currents for both spinless and spin one-half
fermions in a plane pierced by a flux tube (see also \cite{CMO1}), 
the second virial coefficient of interacting anyons, etc., and 
the number of bound states for a flux tube of nonzero radius.
Moreover, in contrast to the \v{S}\v{t}ov\'{\i}\v{c}ek and 
V\'{a}\v{n}a  paper \cite{SV}, the results were applied to discuss
the Hall effect in the presence of magnetic vortices \cite{AM}.

Let $e$, $m$, $E$ be as in \cite{SV}  the charge, the mass,
and the energy of the scattered particle and let us concentrate
on the rotationally invariant case.
Let $\alpha$ be the total flux  through 
the flux tube in units of the flux quantum $\Phi_0=hc/|e|$.
Let us write $\alpha=n+\eta$, where $n$ is an integer and 
$\eta$ is the  nonintegral part of $\alpha$, $0\leq \eta<1$.
Under certain conditions \cite{AM}, one ends up with 
a bound state in either one or in both channels
$l=-n$ and $-n-1$ \cite{AM}. If $E_l$ is the corresponding
bound state energy, the conventional phase shift $\delta_l^0$ 
receives an additional contribution $\triangle_l(E)$ and the resulting
phase shift is
\begin{equation}
\delta_l(E)=\delta_l^0+\triangle_l(E)=
\frac{1}{2}\pi(|l|-|l+\alpha|)+\triangle_l(E),
\label{shift}
\end{equation}
where $\delta_l^0$ corresponds to the conventional 
Aharonov-Bohm scenario and 
\begin{equation}
\triangle_l=
\arctan\left(\frac{\sin(|l+
\alpha|\pi)}{\cos(|l+\alpha|\pi) -A_l^{-1}}\right).
\label{trianl}
\end{equation}
Here $A_{-n} =\left(E/|E_{-n}|\right)^{\eta}$ and 
$A_{-n-1} =\left(E/|E_{-n-1}|\right)^{1-\eta}$
are {\em energy dependent}, and  $A_l\rightarrow 0$ 
as $\eta\rightarrow 0$. In the limit $|E_{l}|\rightarrow 0$,
$\delta_l^0\rightarrow -\delta_l^0$
(the phase-shift flip), while the conventional 
Aharonov-Bohm scenario corresponds to the limit $|E_{l}|\rightarrow \infty$
which yields $\triangle_l=0$. 
Surprisingly enough, the presence of a bound state
manifests itself by a resonance at some positive energy,
which depends on $\eta$ and $|E_l|$ \cite{AM}.
For example, for $0<\eta<1/2$ the {\em resonance} appears at the 
$l=-n$ channel at $E_{res}=|E_{-n}|/[\cos(\eta\pi)]^{1/\eta}>0$.
The differential cross section is then \cite{AM}
\begin{eqnarray}
\lefteqn{
\left(\frac{d\sigma}{d\varphi}\right)(k,\varphi) =
\left(\frac{d\sigma^0}{d\varphi}\right)(k,\varphi)\, +
\frac{8\pi}{k}\sum_{l=-n-1}^{-n}\sin^2\triangle_{l} +
\frac{4}{k}\frac{\sin(\pi\alpha)}{\sin(\varphi/2)}
\times
}\nonumber\\
&&
\left[
\sin\triangle_{-n}\cos\left(\triangle_{-n}-\pi\alpha
+\varphi/2\right) +\sin\triangle_{-n-1}\cos\left(\triangle_{-n-1}
+\pi\alpha
-\varphi/2\right) \right],
\label{dbcross}
\end{eqnarray}
where $k=(2m E/\hbar^2)^{1/2}$. 
The differential cross section remains periodic
with respect to the substitution $\alpha\rightarrow \alpha \pm 1$,
and, as can be easily verified, it becomes {\em asymmetric} 
with regard to  $\varphi\rightarrow-\varphi$ 
(what is equivalent, with regard to
$\alpha\rightarrow -\alpha$) as long as
$\triangle_{-n}\neq -\triangle_{-n-1}$ (modulo $\pi$).
However, as has been discussed in \cite{AM}, this is generic.
The breaking of the rotational symmetry
is  then simply a consequence of the fact that the nonstandard 
boundary conditions can only be imposed in an asymmetric way,
namely for $\alpha\geq 0$ only in the channels $l=-n$ and $-n-1$ 
with $l\leq 0$. 

The asymmetric differential scattering
cross section can give rise to the Hall effect.
Let $n_v$ be the density of vortices and $n_e$  
the density of electrons. The Hall resistivity calculated in 
the dilute vortex limit
(by neglecting multiple-scattering contributions) is \cite{AM} 
\begin{equation}
\rho_{xy}= \frac{4n_v}{n_e}\frac{hc^2}{e^2}\sin(\pi\alpha)\left[
\sin\triangle_{-n} \cos(\triangle_{-n}-\pi\alpha)
+\sin\triangle_{-n-1} \cos(\triangle_{-n-1}+\pi\alpha)\right].
\label{reshall}
\end{equation}
Note that, with  appropriate modifications, results in this paper
apply also to scattering of sound waves in a vortex field \cite{So},
to the case of neutral particles with an 
anomalous magnetic moment in an electric field,
and to exotic cases of scattering in the presence of a cosmic string 
and a gravitational vortex
(see \cite{AM} for references).

This work is part of the research program by  the Stichting voor 
Fundamenteel Onderzoek der Materie  (Foundation for Fundamental Research on
Matter) which  was made possible by financial support 
from the Nederlandse Organisatie voor Wetenschappelijk Onderzoek 
(Netherlands Organization for Scientific Research).


\end{document}